# An enhanced method to compute the similarity between concepts of ontology

Abdelhadi Daoui [1], Noreddine Gherabi [2] and Abderrahim Marzouk [3]

[13] Hassan 1st University, FSTS, IR2M Laboratory, Settat, Morocco
[2] Hassan 1st University, ENSAK, LIPOSI Laboratory, Khouribga, Morocco
{abdo.daoui, gherabi}@gmail.com
amarzouk2004@yahoo.fr

**Abstract.** With the use of ontologies in several domains such as semantic web, information retrieval, artificial intelligence, the concept of similarity measuring has become a very important domain of research. Therefore, in the current paper, we propose our method of similarity measuring which uses the Dijkstra's algorithm to define and compute the shortest path. Then, we use this one to compute the semantic distance between two concepts defined in the same hierarchy of ontology. Afterward, we base on this result to compute the semantic similarity. Finally, we present an experimental comparison between our method and other methods of similarity measuring.

**Keywords:** semantic web, ontologies, similarity measuring, Dijkstra's algorithm.

## 1    Introduction

Today, ontologies play an important role in many domains related to the semantic Web [1], information retrieval [2], knowledge engineering [3] and knowledge management [4]. Therefore, several researches and studies have been developed or are being done to cover this fertile area. These researches can be used in different approaches such as concepts creation, ontology design [5], classification [6], or segmentation [7]. The latter is useful for the processing of large ontologies, which is difficult to maintain, namely the addition, modification or deletion of large ontology parts.

Our work will focus on the measuring of the semantic similarity between concepts of ontology. This one is an important concept used in different areas of research. Jeffrey Hau, William Lee and John Darlington [8] use the semantic similarity to define compatibility between semantic web services [9] [10] annotated by OWL ontologies [11]. In [12] the authors present a method based on multiple information resources (lexical taxonomy, corpus…) to measure the semantic similarity between words. The similarity is also used in the correspondence between the shapes for example, the authors in [13] compute the similarity between outlines of 2D shapes by using a technique based on the extracting of the shapes contours which are represented by a set of points, then the authors describe each segment of this contours by a local and global features, these ones will be coded in string of symbols and stored into XML files On which the similarity calculation will be executed.

Also, several techniques are proposed to compute the semantic similarity between ontologies [14] [15]. Where, the authors, in [15] propose a new method to compute the semantic similarity which is based on three steps. In the first the authors compute the semantic similarity of nodes, and then they compute the semantic similarity of relations between these nodes, at last they combine these two results to form a unified value of semantic similarity.

There are two families of approaches to compute the semantic similarity between concepts:

  1. A family based on computing the geometric distance between concepts to define their semantic similarity, where the less distance gives the more similarity [16].
  2. A family based on degree of information sharing, more common information between two concepts means more similarity [8].

The principal idea of our method is defining the shortest path between any node of a graph (in the current paper the term graph is used to describe ontology) and the root node. Then, we base on these shortest paths and our formula for computing the rate of semantic similarity between the concepts of this graph.

This paper is organized as follows: in Section 2 we describe our method. The next section presents an experimental comparison with some other methods of similarity measuring, followed by a discussion of the changes made to the methodology. Finally, the section 4 presents our conclusion.

## 2    Proposed method

Our method is designed to compute the semantic similarity between two concepts that exist in the same hierarchy of ontology, where all their nodes are connected by "is-a" relations type. This method is summarized in the algorithm shown below in figure 1.

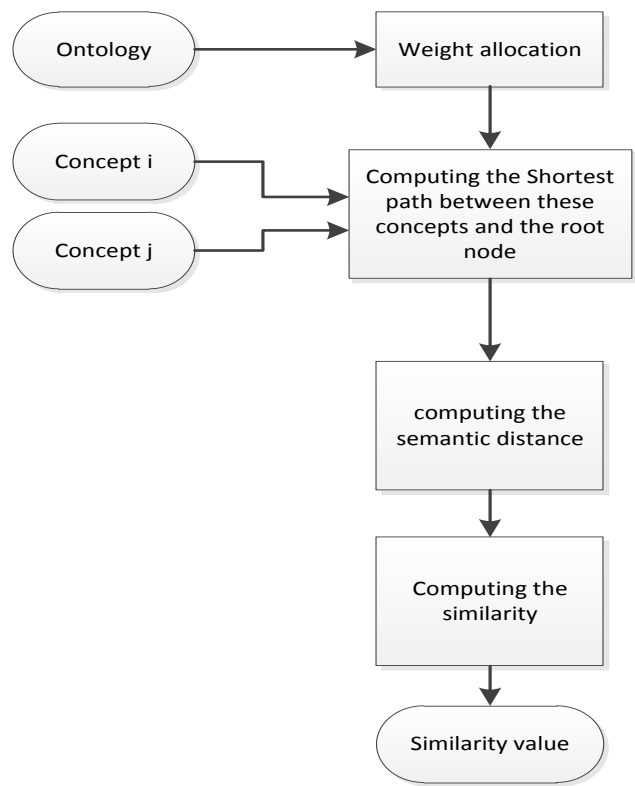

**Fig. 1.** A graphical representation of the proposed algorithm for similarity measuring of ontology concepts.

The first step in our method is reserved for weight allocation to the arcs which represent the relations between the nodes of the studied ontology (section 2.1). Then, we calculate the shortest path from the node, which we want to measure its similarity to the root node (section 2.2). Afterward, section 2.3 is devoted to compute the semantic distance between the two concepts which we need to measure the similarity between them. Then, we use this semantic distance to define the rate of similarity between these two concepts (section 2.4). Finally, in (section 2.5) we present our global algorithm which summarizes our proposed method.

## 2.1 Weight allocation

In the similarity measuring literature, several methods of weight allocation exist, we distinguish between those that affect the value of weight to the nodes like the methods proposed in [17] [18] and the others that allocate the value of weight to the relations (arcs) between nodes [16]. Our method is designed to compute the semantic similarity rate between two concepts of ontology whose all their nodes are connected by the relations of "is-a" type (inheritance type). Therefore, in this paper, we have adjusted one from the second type of weight allocation methods which allows allocating the weight W (m, n) computed using the formula 1 to all the arcs in the ontology.

$$W(\mathrm{m,n}) = [\max(\mathrm{depth}(\mathrm{m})) + \frac{N(n)}{NTNodes(\mathrm{G})+1} + 1]^{-1} \qquad (1)$$

Where, m and n represent two nodes directly connected, max (depth (m)) represents the maximum depth of the node m (the depth of the root node is equal to 0 and 1 for the nodes directly connected to the root node and so on), NTNodes and N (n) represent successively the total Number of nodes in graph G and the order number of the node n between their siblings. And this latter (N (n)) is an integer greater or equal to 0.

We consider the following ontology:

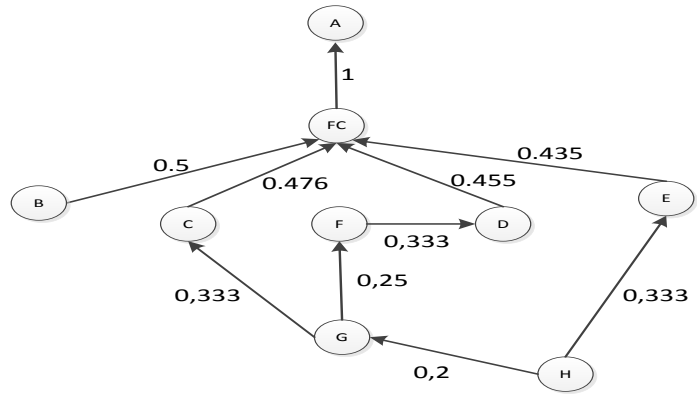

**Fig. 2.** An example of ontology.

The weights of arcs mentioned in the figure 2 are calculated by our formula (formula 1) of weight allocation. In this formula, we have taked max (depth (m)) for ensuring that the weight of the current arc is always less than the weights of their previous arcs. Therefore, the semantic similarity between two concepts more specific is greater than two concepts more generalist.

## 2.2    Shortest path defining

The shortest path is a famous problem in the science research domain. Therefore, several algorithms are proposed such as Dijkstra's algorithm [19], Bellman-Ford algorithm [20], Floyd-Warshall algorithm [21], but each one of these is designed to define the shortest path under some criteria. In this paper, with our criteria which are: (1) the weight is strictly greater than zero. (2) The relations between the nodes are "is-a" type.

The best solution adapted to our case is Dijkstra's algorithm with the use of the formulas (2 and 3). For this, we have adapted this one with some modifications that allow stopping the algorithm once reaches the root node, which is beneficial in the performance or in the execution time. From our second criterion (The relations between the nodes are "is-a" type), we can deduce that all the arcs of the graph will be oriented to the root node. in the calculation process of the shortest path, from a given node to the root node, only the nodes that can be considered a generalization of the current node will be visited and not all the nodes of the graph.

$$W_0[m,n] = \begin{cases} 0 \; if \; m = n \\ \infty \; m \neq n \end{cases} \quad (2)$$

*For all* $0 \leq k \leq S - 1$

$$W_{k+1}[m,n] = \min \begin{cases} W_k[m,n] \\ W_{k+1}[m,x] + G[x,n] \end{cases} \quad (3)$$

m, n and x represent three nodes, S represents the set of all nodes of graph. The nodes x and n are directly connected and $W_k$ [m, n] represents the weight of path [m, n] at iteration k,

In our adapted algorithm, for defining the shortest path, we use the formula 2 for initializing the nodes weight, where we give the value zero to the start node and infinity to all other nodes in the graph, and we use the formula 3 to compute the weight of the shortest path which exists between the start node and the root node.

Our adapted algorithm is designed as follows:

```
Function shortestPath(G,node,rootNode)
F=S   // S represents a set of all nodes of graph G
for t in S do // t represents a node in S
  If t == node then
    W[t]=0
  Else
    W[t]= ∞
  End if
  π [t]= ∅ //π [t] represents the antecedent node of
//node t
end for
while F not empty  do
  t=Node extracted from F with smallest Weight
  for s successor of t do
    if  W[t]+W[t,s]<W[s] then
      W[s] =  W[t]+ W[t,s]
      π [s] = t
      if s == rootNode then
        return shortestPath
      end if
    end if
  end for
  delete t from F
end while
```

**Fig. 3.** Our adapted algorithm for defining the shortest path.

This algorithm defines the shortest path between a given node in the ontology and the root node. In contrast to the Dijkstra's algorithm, this algorithm will stop the execution once the root node is reached.

In addition, our algorithm allows not only defining the shortest path, but also defining the value of its weight. And we can't define the shortest path without defining its weight.

For example, we consider the ontology presented in the figure 2. To define the shortest path from the node H. There are three paths from H to A (the root node): (H, E, FC, A), (H, G, F, D, FC, A) and (H, G, C, FC, A).

**Table 1.** An example to define the weight of shortest path.

|         | A     | FC    | B        | C     | D     | E     | F     | G     | H |
|---------|-------|-------|----------|-------|-------|-------|-------|-------|---|
| $W_0$   | $\infty$ | $\infty$ | $\infty$ | $\infty$ | $\infty$ | $\infty$ | $\infty$ | $\infty$ | 0 |
| $W_1$   | $\infty$ | $\infty$ | $\infty$ | $\infty$ | $\infty$ | 0.333 | $\infty$ | 0.2 | 0 |
| $W_2$   | $\infty$ | $\infty$ | $\infty$ | 0.533 | $\infty$ | 0.333 | 0.45 | 0.2 | 0 |
| $W_3$   | $\infty$ | 0.768 | $\infty$ | 0.533 | $\infty$ | 0.333 | 0.45 | 0.2 | 0 |
| $W_4$   | $\infty$ | 0.768 | $\infty$ | 0.533 | 0.783 | 0.333 | 0.45 | 0.2 | 0 |
| $W_5$   | $\infty$ | 0.768 | $\infty$ | 0.533 | 0.783 | 0.333 | 0.45 | 0.2 | 0 |
| $W_6$   | 1.768 | 0.768 | $\infty$ | 0.533 | 0.783 | 0.333 | 0.45 | 0.2 | 0 |

➡ Exit

We can easily observe in the table 1, that the shortest path between the node H and the root node is equal to 1.768.

**Table 2.** An example to define the shortest path.

|         | A  | FC | B   | C  | D   | E  | F  | G  | H   |
|---------|----|----|-----|----|-----|----|----|----|-----|
| $W_0$   | ∅  | ∅  | ∅   | ∅  | ∅   | ∅  | ∅  | ∅  | ∅   |
| $W_1$   | ∅  | ∅  | ∅   | ∅  | ∅   | H  | ∅  | H  | ∅   |
| $W_2$   | ∅  | ∅  | ∅   | G  | ∅   | H  | G  | H  | ∅   |
| $W_3$   | ∅  | E  | ∅   | G  | ∅   | H  | G  | H  | ∅   |
| $W_4$   | ∅  | E  | ∅   | G  | F   | H  | G  | H  | ∅   |
| $W_5$   | ∅  | E  | ∅   | G  | F   | H  | G  | H  | ∅   |
| $W_6$   | FC | E  | ∅   | G  | F   | H  | G  | H  | ∅   |

From the table 2, we can observe that:

$\pi$ [A]  = FC
$\pi$ [FC] = E ⎫
$\pi$ [E]  = H ⎭ ➡ SPath (H, E, FC, A)

Where, SPath (H, E, FC, A) represents the shortest path between the nodes H and A defined by our adapted algorithm.

## 2.3 Semantic distance computing

At this level, to compute the semantic distance between two concepts we use a new technique based on the shortest path calculated in the previous phase. This technique allows eliminating unnecessary parts in the shortest paths and only keeps the necessary parts to calculate the semantic distance between two concepts.

This technique is designed as follows:

We consider the colored segment in the figure below which represents a segment of the ontology mentioned in figure 2.

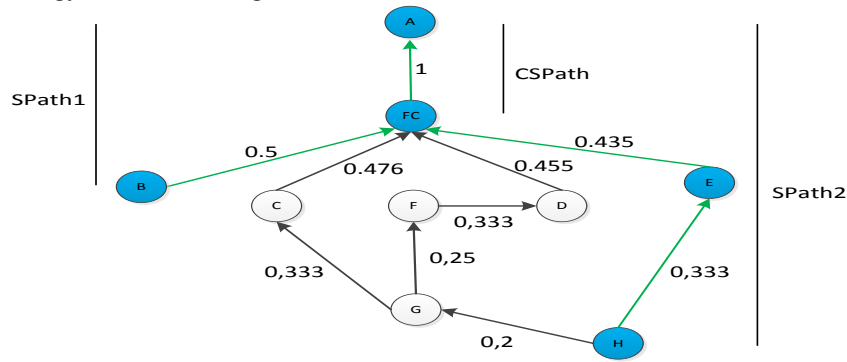

**Fig. 4.** A segment of the ontology mentioned previously.

Where, SPath1 represents the shortest path between the nodes B and A. The SParh2 represents the shortest path between the nodes H and A. FC represents the first common node between SPath1 and Spath2 coming from H and B. And CSPath represents common sub shortest path between the FC and the root node.

For compute the semantic distance between any two concepts exist in the same graph, we use the formula 4.

$$\text{SDis}(C1, C2) = W[\text{SPath1}] + W[\text{SPath2}] - 2 * W[\text{CSPath}] \tag{4}$$

Where, C1 and C2 represent two concepts (in the figure 4 C1 represents the concept (B, FC, A) and C2 represents the concept (H, E, FC, A)) and W[SPath$_i$] is defined as follows:

$$W\left[SPath_i\right] = \sum_{j=1}^{k} W_j[m, n] \tag{5}$$

m and n represent two nodes directly connected in SPath$_i$ and k represents the set of arcs in SPath$_i$.

### 2.4 Semantic similarity computing

There is an inverse relation between semantic similarity and semantic distance. Increasing of one among them decrease the other. We can categorize the output of the semantic similarity function in three categories:

1. The two concepts are the same.
2. Nothing in common between them.
3. There is a rate of semantic similarity between them.

Therefore, the function of semantic similarity should verify these three conditions:

1. $\forall\,(C1,C2) \in G : 0 \leq SSim(C1,C2) \leq 1$
2. $\forall\,C1 \in G : SSim(C1,C1) = 1$
3. $\forall\,(C1,C2,C3) \in G : if\;SDis(C1,C2) > SDis(C1,C3)\;then$
   $\quad SSim(C1,C2) < SSim(C1,C3)$

Where, SSim represents the semantic similarity, SDis represents the semantic distance and (C1, C2 and C3) represent three concepts of graph G. Ci represents the set of nodes constituting the shortest path between a given node and the root node.

In this paper, we have used the function of semantic similarity computing proposed in [16].

$$SSim(C1,C2) = \frac{1}{deg*\,SDis(C1,C2) + 1} \tag{6}$$

C1 and C2 represent two concepts and the parameter " deg " represents the impact degree of Semantic distance on semantic similarity, and it should be between $0 < deg \leq 1$ (the concrete value of "deg" is defined in the experience).

### 2.5 Global algorithm

This section is devoted to our global algorithm that includes all the previous algorithms and techniques.

```
Input: C1,C2,rootNode,Graph
Output: semantic similarity value
If C1==C2 then
     SDis=0
Else if C1 is directly connected to C2 then
     SDis=W[C1,C2]
Else
     SPath1 =ShortestPath(Graph,C1,rootNode)
     SPath2 =ShortestPath(Graph,C2,rootNode)
     SDis = W [SPath1] + W [SPath2] − 2*W [CSPath]
End if
```
$$SSim = \frac{1}{deg*\,SDis + 1}$$
```
Return SSim
```

**Fig. 5.** Our global algorithm.

## 3 Experiments

In this section, by consulting the WordNet, we have got a fragment of ontology hierarchy shown in Figure 6, concerning the terms: Vehicle, truck, car, family car, sport car, luxury car and bus which we want to compare the similarity between them.

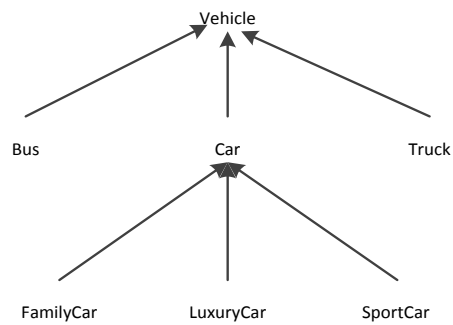

**Fig. 6.** A fragment of ontology hierarchy.

After computation of semantic similarity between these concepts by using our method (where, we have fixed the parameters deg to 0,4) and two others, we have obtained the results presented in the following tables.

In this experimental comparison, each node presents in the following tables such as Truck, Car and so on, represents a concept which is constituted from the set of all nodes of the shortest path which exists between this node and the root node. For example the node "SportCar" presents in these tables represents the concept (SportCar, Car, Vehicle).

**Table 3.** Computing the semantic similarity using the method [22].

|  | Vehicle | Truck | Car | Family Car | Sport Car |
|---|---|---|---|---|---|
| **Vehicle** | 1 | 0 | 0 | 0 | 0 |
| **Truck** | 0 | 1 | 0 | 0 | 0 |
| **Car** | 0 | 0 | 1 | 0 | 0 |
| **Family Car** | 0 | 0 | 0 | 1 | 0 |
| **Sport Car** | 0 | 0 | 0 | 0 | 1 |

**Table 4.** Computing the semantic similarity using the method [16].

|  | Vehicle | Truck | Car | Family Car | Sport Car |
|---|---|---|---|---|---|
| **Vehicle** | 1 | 0.71 | 0.71 | 0.58 | 0.58 |
| **Truck** | 0.71 | 1 | 0.55 | 0.47 | 0.47 |
| **Car** | 0.71 | 0.55 | 1 | 0.76 | 0.76 |
| **Family Car** | 0.58 | 0.47 | 0.76 | 1 | 0.62 |
| **Sport Car** | 0.58 | 0.47 | 0.76 | 0.62 | 1 |

**Table 5.** Computing the semantic similarity using our method.

|  | Vehicle | Truck | Car | Family Car | Sport Car |
|---|---|---|---|---|---|
| **Vehicle** | 1 | 0.758 | 0.738 | 0.643 | 0.652 |
| **Truck** | 0.758 | 1 | 0.597 | 0.533 | 0.54 |
| **Car** | 0.738 | 0.597 | 1 | 0.833 | 0.849 |
| **Family Car** | 0.643 | 0.533 | 0.833 | 1 | 0.726 |
| **Sport Car** | 0.652 | 0.54 | 0.849 | 0.726 | 1 |

By analyzing of the obtained results of semantic similarity, we observe that the first method [22] can only find the total similarity (otherwise, it can only discover the similarity between the same concepts). The second method [16] can find the semantic similarity between concepts but with a low score. Finally, our method can also calculate the semantic similarity between concepts but with a high score.

The second method [16] allows affecting the same value of the semantic similarity between the sibling concepts and their parent concept (for example, the semantic similarity between the concepts (Vehicle) and (Vehicle, Truck) is equal to 0.71 and the same value of semantic similarity between (Vehicle) and (Vehicle, Car)), but in the reality the rate of the semantic similarity between these concepts differs. For this reason, we have proposed our method which allows avoiding this type of problems by affecting the different values of semantic similarity for similar cases.

Also our method uses a simple function of weight allocation and bases on famous and robust algorithms with some adaptations to minimize the time of execution. Therefore, this one can calculate the semantic similarity in very short time even in industrial size ontologies.

## 4    Conclusion

In this paper, we have presented a new method to calculate the similarity between two concepts of ontology. Then, we have compared it against other methods that already exist. The obtained results are very interesting and prove the strength of our method.

We are interested in the future works to adjust this method to support the calculation of similarity between any two concepts that may exist in the same ontology or not and be related by any type of relations.